\documentclass[9pt,twocolumn,twoside]{pnas-new}

\templatetype{pnasresearcharticle} 

\usepackage{soul}

\begin{document}

\title{A Size-Dependent Ideal Solution Model for Liquid-Solid Phase Equilibria Prediction in Aqueous Organic Solutions}

\author[a]{Spencer P. Alliston}
\author[a]{Chris Dames}
\author[b,c,d,1]{Matthew J. Powell-Palm}

\affil[a]{Department of Mechanical Engineering, University of California, Berkeley, Berkeley, CA  94720}
\affil[b]{J. Mike Walker '66 Department of Mechanical Engineering, Texas A\&M University, 400 Bizzell St, College Station, TX  77840}
\affil[c]{Department of Materials Science and Engineering, Texas A\&M University, 400 Bizzell St, College Station, TX  77840}
\affil[d]{Department of Biomedical Engineering, Texas A\&M UNiversity, 400 Bizzell St, College Station, TX  77840}

\leadauthor{Alliston}

\significancestatement{The ideal solution model provides the conceptual foundation of solution thermodynamics, and a useful tool for a priori phase equilibria prediction based only on pure component properties. However, solutions of water and organic molecules tend to deviate strongly from ideality, a phenomenon which is frequently attributed to hydrogen bonding. Here, we demonstrate that much of this deviation can instead be attributed to the effect of differing molecule sizes on the entropy of mixing, and that accounting for these sizes can significantly enhance the accuracy of the ideal model. The size-dependent ideal solution model derived herein provides increased predictive power over liquid-solid equilibria in aqueous organic solutions, and offers new insight into the outsize role of entropy in these systems.}

\authorcontributions{Please provide details of author contributions here.}
\authordeclaration{Please declare any competing interests here.}

\correspondingauthor{\textsuperscript{1}To whom correspondence should be addressed.\\ E-mail: powellpalm\@tamu.edu}


\begin{abstract}
Predictive synthesis of aqueous organic solutions with desired liquid-solid phase equilibria could drive progress in industrial chemistry, cryopreservation, and beyond, but is limited by the predictive power of current solution thermodynamics models. In particular, few analytical models enable accurate liquidus and eutectic prediction based only on bulk thermodynamic properties of the pure components, requiring instead either direct measurement or costly simulation of solution properties. In this work, we demonstrate that a simple modification to the canonical ideal solution theory accounting for the entopic effects of dissimilar molecule sizes can transform its predictive power, while offering new insight into the thermodynamic nature of aqueous organic solutions. Incorporating a Flory-style entropy of mixing term that includes both the mole and volume fractions of each component, we derive size-dependent equations for the ideal chemical potential and liquidus temperature, and use them to predict the binary phase diagrams of water and 10 organic solutes of varying sizes. We show that size-dependent prediction outperforms the ideal model in all cases, reducing average error in the predicted liquidus temperature by 59\%, eutectic temperature by 45\%, and eutectic composition by 43\%, as compared to experimental data. Furthermore, by retaining the ideal assumption that the enthalpy of mixing is zero, we demonstrate that for aqueous organic solutions, much of the deviation from ideality that is typically attributed to molecular interactions may in fact be explained by simple entropic size effects. These results suggest an underappreciated dominance of mixing entropy in these solutions, and provide a simple approach to predicting their phase equilibria.
\end{abstract}

\dates{This manuscript was compiled on \today}
\doi{\url{www.pnas.org/cgi/doi/10.1073/pnas.XXXXXXXXXX}}

\maketitle
\thispagestyle{firststyle}
\ifthenelse{\boolean{shortarticle}}{\ifthenelse{\boolean{singlecolumn}}{\abscontentformatted}{\abscontent}}{}

\firstpage{3}

\dropcap{U}nderstanding the phase equilibria of liquid solutions is essential to many aspects of modern industry, from industrial chemistry to medicine and beyond. However, determination of even simple liquid-solid phase diagrams often requires direct synthesis and measurement of the solution of interest, due to the limited predictive power of current solution thermodynamics models. For many systems, the predictive standard for these phase diagrams remains the Ideal Solution Model, a simplified physical theory dating back more than a century to Raoult and Gibbs. While this model represents certain systems quite well \cite{Martins2019}, the liquid-solid equilibria of many aqueous organic solutions have proven challenging to predict \textit{a priori} (using only properties of the pure components), slowing exploration of the chemical parameter space available to cryopreservation \cite{Murray2022}, natural deep eutectic solvents \cite{Paiva2013}, drug discovery \cite{Garbett2012, Klebe2015}, and others. Building on several foundational works from the last century, we address this canonical problem with a simple modification to the ideal solution model, which both substantially improves predictive power and grants renewed insight into the nature of liquid solution thermodynamics. 

The classical ideal solution model relies on the core assumptions that all molecules in a solution interact in the same manner, and that all molecules contribute equally to the entropy of mixing (implying that all molecules are treated as fungible for the calculation of mixing parameters). In thermodynamic terms, the ideal model requires that the enthalpy of mixing in solution $\Delta H_{mix} = 0$, and that the entropy of mixing $\Delta S_{mix} = -R \sum x_i \ln x_i $, where R is the universal gas constant and $x_i$ is the mole fraction of each component of the solution. These assumptions generally hold well for dilute solutions, and can also prove tenable at higher concentrations in non-contact regimes (as in an ideal gas) or when molecules are constrained to ordered local sites (as in a crystal lattice) \cite{oates1969ideal}. However, in many other systems, and particularly in liquids, one or both of these assumptions may fail, leading to significant divergences from predicted behavior \cite{Hildebrand1927, HildebrandBook}.

It is often presumed that experimental deviations from ideality are a product of unaccounted for interactions between molecules, which may yield a non-zero enthalpy of mixing, as seen in the widely used regular solution model \cite{Gordon1968, robinson1949tables}, or otherwise affect the solution entropy \cite{arrhenius1912theories, Martin2023}. Here, we demonstrate that for many aqueous organic solutions, this deviation from ideal solution behavior is not in fact dominated by the contribution of differing molecular interactions, but instead by the contribution of differing molecule sizes to the entropy of mixing. 

This contribution was first recognized by Flory and Huggins in seminal works on polymer thermodynamics, in which each had the insight that configurational entropy in liquids must depend on the size (and orientation) of the polymer chains present \cite{Flory, Huggins1942}. Here, we adopt a simplified form of the Flory entropy of mixing \cite{Flory, Powell-Palm2024}, which accounts for molecule size alone, and is therefore of more general interest for the study of non-poylmeric molecular solutions:

\begin{equation}
    \Delta S_{\text{mix}} = -R \sum x_i \ln \phi_i
\end{equation}

This relation incorporates the effects of molecular size by way of the volume fractions $\phi_i$ of the components \textit{i} in solution, as compared to the classic ideal equivalent, which relies only on the mole fractions $x_i$. The theoretical applicability of such an entropy descriptor to non-polymer solutions with molecules of differing size was demonstrated by Hildebrand in 1948, but he and others suggested then \cite{Hildebrand1948} and in subsequent works \cite{HildebrandBook} that differences from the ideal model were negligible in regimes of practical interest.  Indeed, for solutions of molecules with relatively similar molar volumes (ratios of less than 2:1) -- including many liquid metals and model solvent solutions -- deviation from the classic ideal model is small.

However, organic solutions that include a small-molar volume solvent (such as water, at 18 mL/mol in standard conditions) and larger solute molecules (such as alcohols, sugars, acids, biomolecules, methylated compounds, etc.) routinely see volume differences much larger than this -- with molar volume ratios regularly exceeding 4:1. In this case, the entropic effect of molecule size becomes increasingly critical to understanding (and predicting) system behavior. This insight has recently been studied in the context of nucleation kinetics in aqueous systems \cite{Powell-Palm2024}, and has been applied to a limited extent in the study of hydrocarbons \cite{coutinho1995, Hansen1988}, but its application to condensed phase equilibria in aqueous organic systems remains, to our knowledge, unexplored.

Here, we seek to develop a simple but general size-dependent theory of the thermodynamics of aqueous organic solutions. We begin by re-deriving essential aspects of the classical ideal solution model (the chemical potential and the Gibbs-Helmholtz liquidus temperature relation), but now incorporating the size-dependent entropy of mixing given in Eqn. (1). Comparing to experimental data from the literature for 10 aqueous binary solutions, we then demonstrate that this simple modification significantly improves the model's power to predict binary phase diagrams, including eutectic temperatures and compositions. Critically, we accomplish this without compromising the key value proposition of the ideal solution model: prediction based solely on the properties of the pure components. 

In addition to improvement of simple thermodynamic predictions, this size-dependent model reveals that mixing entropy has a stronger role in the thermodynamics of aqueous organic solutions than widely appreciated, an insight which may have meaningful implications in predictive synthesis and materials discovery. We anticipate that this theory may prove particularly relevant to the search for "natural deep eutectic solvents” (NADES) \cite{Paiva2013, Hubel2021} and new solutions for cell, tissue, and organ cryopreservation \cite{Murray2022, Ahmadkhani2024}, which often rely upon the large organic molecules examined here. Throughout this work, we restrict our analysis to non-ionic organic systems, for which description of molecule sizes in solution is straightforward (i.e. excess solution volumes are minimal).

\section*{The Size-Dependent Ideal Solution Model}

We begin by incorporating Eqn. (1) into the classic formulation of the molar Gibbs free energy of an ideal solution, which includes a mechanical mixture of the Gibbs free energies of the pure components $G_i$ plus the entropy of mixing:

\begin{equation}
    G_{\text{soln}} = \sum G_i x_i + RT \sum x_i \ln \phi_i
\end{equation}

Here, $x_i$ and $\phi_i$ are the mole fractions and volume fractions of each component i, where $\phi_i = x_i v_i / \sum x_i v_i$, $v_{i}$ is the molar volume of component i, and $T$ is the absolute temperature of the solution. 

This equation can be differentiated with respect to $x_A$, the mole fraction of component $A$, and combined with standard thermodynamic relations to recover the chemical potential $\mu_A$ \cite{porter} of component $A$ in solution, shown here for a binary solution of components $A$ and $B$:

\begin{equation}
    \mu_A = G_A + RT \left[ \ln (\phi_A) + \phi_B - \frac{x_B}{x_A} \phi_A  \right]
\end{equation}

where $G_A$ is the molar Gibbs free energy of component $A$ in its pure form. Hansen  \cite{Hansen1988}, Coutinho \cite{coutinho1995}, Hildebrand \cite{HildebrandBook}, and colleagues each arrived at similar forms of this chemical potential in their study of hydrocarbons; our subsequent derivations (and application to aqueous organic solutions) have, to our knowledge, not been reported previously. 

Given that for a binary system $x_A + x_B = 1$ and $\phi_A + \phi_B = 1$, and defining a component molar volume ratio $V_{R,A} = v_{B}/v_{A}$, it can be shown that the chemical potential of component $A$ is a function only of $\phi_A$ and $V_R$.

\begin{equation}
    \mu_A = G_A + RT \left[ \ln (\phi_A) + (1-\phi_A) \left( 1-\frac{1}{V_{R,A}} \right) \right]
\end{equation}

This chemical potential is the practical foundation of the size-dependent ideal solution theory. Notably, this equation relies only on the assumption that the Gibbs free energies of the pure components do not change in solution. Therefore, this result is highly general, and is suited to serve as the basis for an expanded size-dependent thermodynamic framework.

From this equation, there are multiple approaches from which to predict liquidus behavior. Given our desire for a predictive model requiring only select properties of the pure components, we chose to predict phase equilibria using a Gibbs-Helmholtz analysis. We assume first that each component forms a pure solid (and therefore has no mixing entropy in the solid phase). This is a reliable assumption for water, which forms a nearly pure crystal in the ice Ih phase \cite{Murray2008, Hudait2014}, and holds generally for the crystalline forms of many large organic compounds \cite{Sangster1999}. Furthermore, the assumption is applied in both the classic and size-dependent models, and should therefore not confound comparison. We also assume that the system exists under isobaric conditions.

Following the standard Gibbs-Helmholtz derivation of liquidus temperature \cite{Martin2023}, we equate the chemical potential of component $A$ in the solid phase, which is simply $G_{A,s}$, with that in the liquid phase, which takes the form of Eqn. (4). This relation defines the equilibrium point and leads to

\begin{equation}
    G_{A,s}  = G_{A,l} + RT \left[ \ln (\phi_A) + (1- \phi_A) \left( 1- \frac{1}{V_{R,A}} \right) \right]
\end{equation}

which can be rearranged to isolate the volume fraction:

\begin{equation}
     \ln (\phi_A) + (1- \phi_A) \left( 1- \frac{1}{V_{R,A}} \right) = \frac{G_{A,s} - G_{A,l}}{RT} = - \frac{ \Delta_{fus} G_A}{RT}
\end{equation}

where $\Delta_{fus} G_A$ is the Gibbs free energy of fusion, defined as $G_{A,l}-G_{A,s}$
Next, we differentiate the above with respect to temperature, and introduce the enthalpy of fusion, $\Delta_{fus} H_A$ of the pure component using the Gibbs-Helmholtz equation, which is valid under isobaric conditions:

\begin{equation}
    \frac{d}{d T} \left[ \ln (\phi_A) + (1- \phi_A) \left( 1- \frac{1}{V_{R,A}} \right) \right] = \frac{ \Delta_{fus} H_A^0}{RT^2}
\end{equation}

From here, we integrate from the limits of $T_{liq,A}$, the liquidus temperature of the solution and solid component $A$, to $T_{m,A}$, the melting point of pure component $A$. Note that, for the left side of the equation, the composition can be treated as a function of temperature, with  $\phi_A(T_{liq,A}) = \phi_A$ and $\phi_A(T_{m,A}) = 1$, and an analogous relation can be produced for $x_A$. We also assume that the enthalpy of fusion does not vary significantly with temperature, such that $\Delta_{fus} H_i(T) = \Delta_{fus} H_i^0$, the enthalpy of fusion at the melting temperature, which holds well for relatively small temperature differences \cite{Martins2019, coutinho1995}. This yields

\begin{eqnarray}
    \ln \left(\frac{1}{\phi_A} \right) - (1- \phi_A) \left( 1- \frac{1}{V_{R,A}} \right) \nonumber \\ = - \frac{ \Delta_{fus} H_A^0}{R} \left( \frac{1}{T_{m,A}} - \frac{1}{T_{liq,A}} \right)
\end{eqnarray}

Equation (8) can finally be solved for the liquidus temperature as a function of concentration:

\begin{equation}
    T_{liq,A} = \left( \frac{1}{T_{m,A}} - \frac{R [ \ln (1/\phi_A) - (1-\phi_A ) (1-1/V_{R,A})]}{\Delta_{fus}H^0_A} \right)^{-1}
\end{equation}

Eqn. (9) shows that the only material data needed to predict binary liquidus curves using the size-dependent ideal solution model are the molar volume, enthalpy of fusion, and melting temperature of the pure components, all of which can be readily obtained from the literature for most materials. The model does not depend on any experimental information from the solution itself, and therefore retains the predictive utility of the classic ideal solution model. 

Within this framework, any $V_{R,A} \neq 1$ will lower the liquidus temperatures of both components relative to the ideal prediction for a given mole fraction. We note that while this is generally consistent with physical observation (and especially for aqueous systems),  Martin and Shipman have recently identified select heavily-interacting systems with eutectic temperatures higher than ideal prediction\cite{Martin2023}. Additionally, portions of 2 experimental liquidus curves in this work (sorbitol, maltitol) are higher than predicted by the ideal model, even while their eutectic temperatures are depressed.   As expected, the classic formulation of the eqn. (9) is recovered when $V_{R,A} = 1$.

\begin{figure}[h!]
\centering
\includegraphics[width = .85\linewidth]{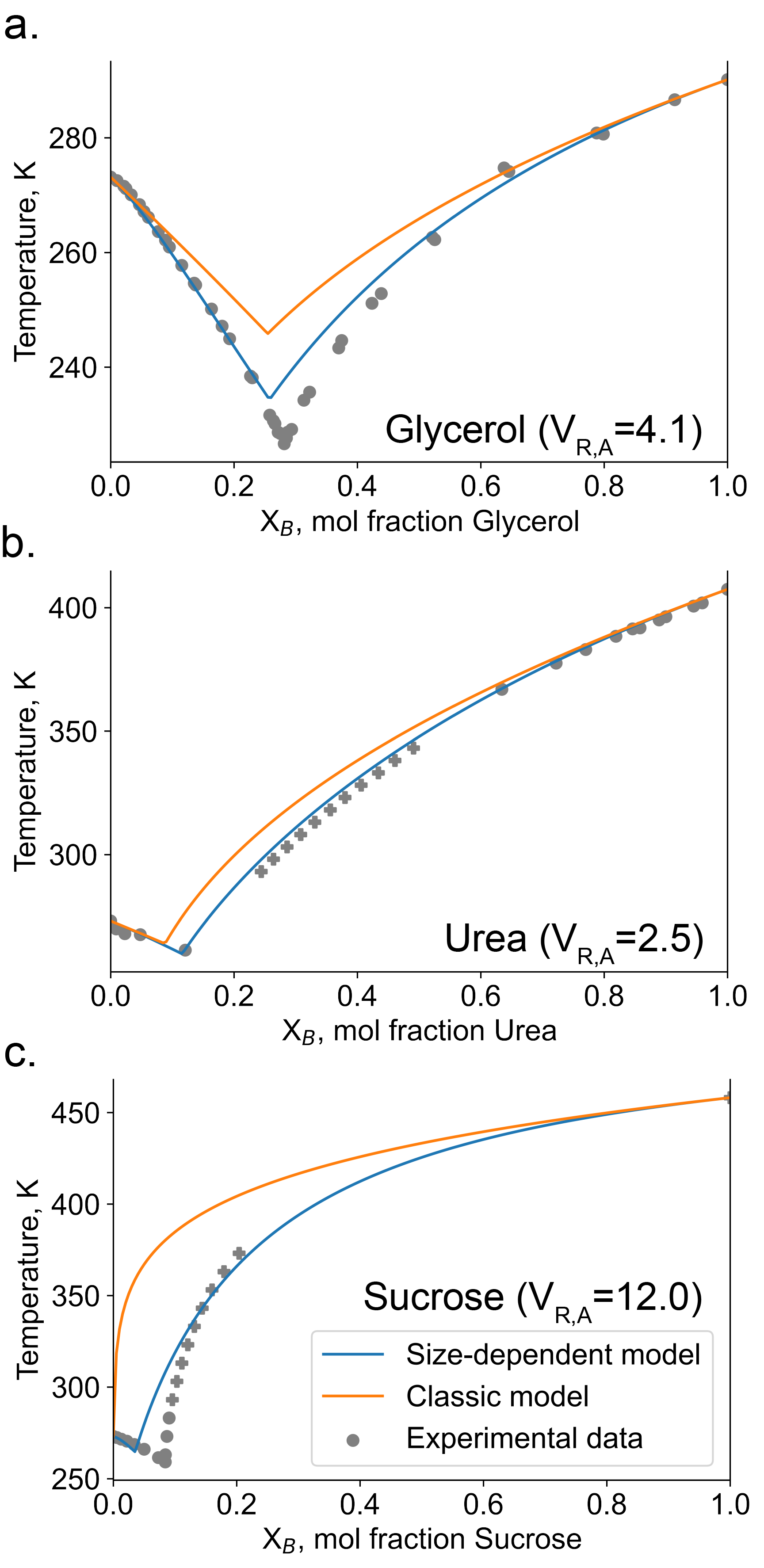}
\caption{\label{fig:VIMEx} Example phase diagram predictions for representative aqueous organic solutions. (a) Classic ideal prediction vs. size-dependent prediction for glycerol and water, a solution frequently used in cryopreservation. Analogous data is shown for (b) urea and water and (c) sucrose and water, which are the smallest and largest solutes included in this study. Urea and sucrose have each previously had their depressed liquidus curves attributed to hydrogen bonding \cite{Martin2023}. It can be seen that the prediction is improved in both cases, but that the effect in urea can be almost entirely attributed to volumetric factors while sucrose retains aspects not captured by the size-dependent scheme. Similar phase diagrams can be seen for all sampled solutions in the Supplemental Information (Figure S1).}
\end{figure}

\begin{figure*}[h!]
\includegraphics[width = 17.4 cm]{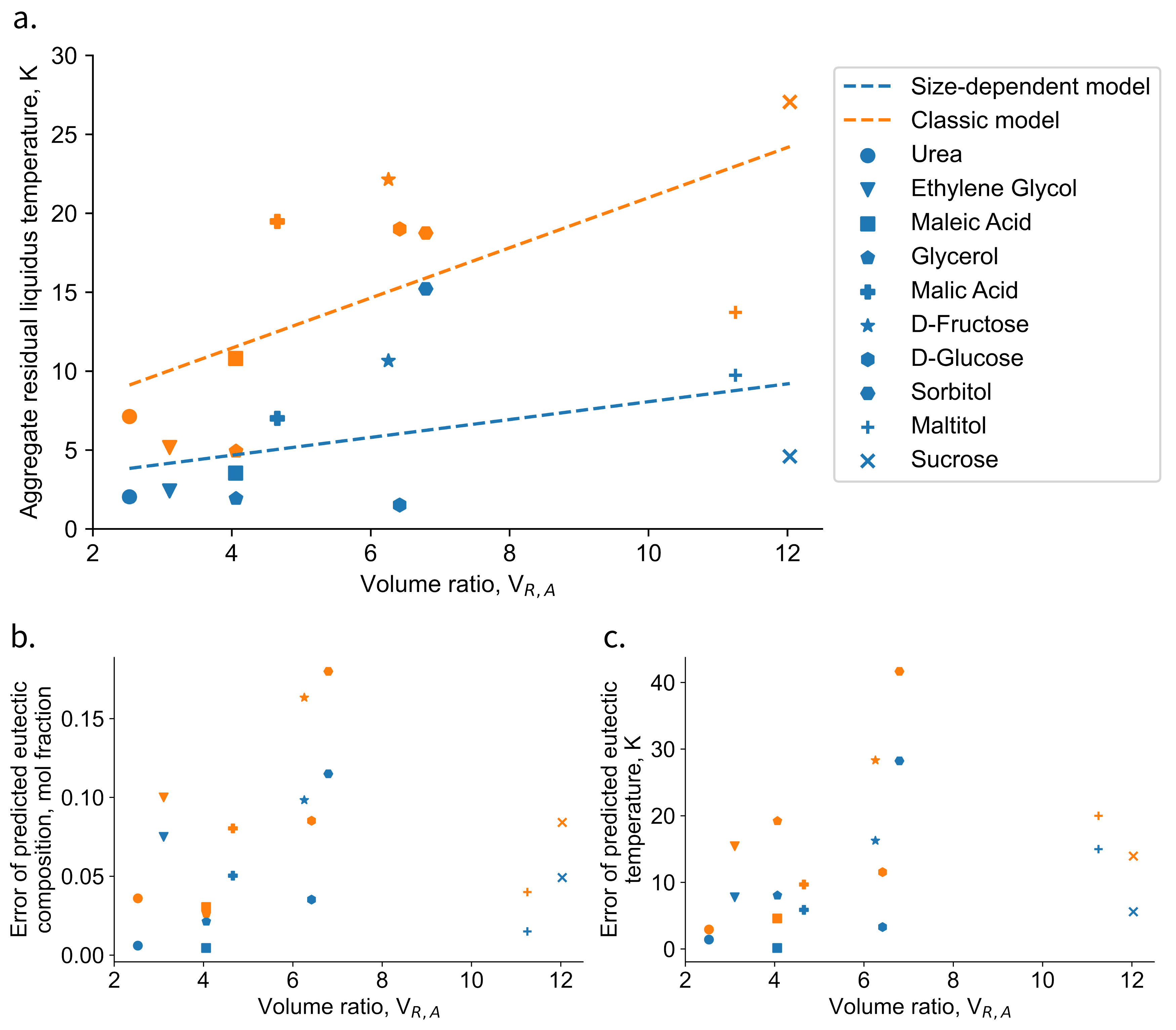}
\caption{\label{fig:VIMRes} (a) Residual temperatures of predicted liquidus curves. Data shown are trapezoidal integrations of the difference between literature experimental data and the model curves. (b) Error in predicted composition of eutectic point. (c) Error in predicted eutectic temperature. Data for (b) and (c) are shown in absolute difference from experimentally-determined eutectic point in mole fraction and K, respectively. Predictions using the size-dependent model are improved in all cases. Experimental equilibria were sourced for Urea from Babkina and Kuznetsov \cite{Babkina2010} and Shnidman and Sunier \cite{Shnidman} (+ in Fig. 1.b), Ethylene Glycol (metastable) from Ott et al. \cite{Ott1972}, Maleic Acid and Malic Acid from Beyer et al. \cite{Beyer2008}, Glycerol from Lane \cite{Lane1925}, D-Fructose (metastable) from Young \cite{Young1952}, $\alpha$-D-Glucose from Young \cite{Young1922}, Sorbitol and Maltitol from Siniti et al \cite{Siniti1999}., and Sucrose from Young and Jones \cite{Young1949} and Stephen and Stephen \cite{Stephens} (+ in Fig. 1.c).}
\end{figure*}

By calculating the liquidus temperature for each of the constituent molecules across the entire concentration range (and choosing $V_R$ accordingly to be with respect to each constituent as a solute) and retaining only the higher, stable liquidus at a given concentration, the equilibrium phase diagram for the solution can be constructed.

\section*{Comparison of model to experimental data}

In order to validate the predictions of the proposed model and facilitate comparison to the classic ideal model, we collected 10 published phase equilibria datasets from the literature across a range of $V_R$. We restricted our analysis to aqueous organic systems exhibiting simple eutectic behavior (i.e. including only a eutectic point between two pure solids and the liquid), and for which liquidus data on both sides of the eutectic were available. No ionic solutes were used, as interpretation of dissociated ion molar volumes is not straightforward, and ionic molecular interactions can produce extreme excess volumes in solution. For ethylene glycol and d-fructose, experimental metastable simple eutectics were used. Maltitol and sorbitol had experimental eutectic temperatures, but approximate eutectic compositions were interpolated for these solutions.

Inputs were limited to properties of the pure components, namely their enthalpy of fusion values, molar volumes at room temperature, and melting temperatures. All of these were sourced from the NIST Chemistry Webbook unless otherwise specified. Tabulated source data can be found in the Supplemental Information (Table S1) of this work. Phase diagrams of liquidus temperatures vs. mole fraction were calculated using these parameters and equation (9).

Figure~\ref{fig:VIMEx} shows representative datasets used to evaluate the ability of this model to predict liquidus behavior. Figure~\ref{fig:VIMEx}.a shows the water-glycerol phase diagram, which has a $V_{R,A}$ = 4.1 and is a solution of interest in cryobiology. It can be seen that the experimental eutectic is significantly depressed relative to the classic model, but is well-predicted by the size-dependent ideal model. The remaining discrepancy between prediction and experiment likely stems from the idealized assumptions of the present model, especially that of constant enthalpy of fusion, which has been shown to become less robust at temperatures distant from the pure phase melting points (such as near the eutectic temperature)\cite{Martin2023}.

Figure~\ref{fig:VIMEx}.b and .c show the phase diagrams of water-urea and water-sucrose, two solutions that have divergences from ideality that have previously been attributed to molecular interactions (specifically, hydrogen bonding) \cite{Martin2023}. Here, we find that most of the divergence of aqueous urea can be attributed to simple entropic size effects as captured in Eqn. (9), while aqueous sucrose appears to be affected significantly by both entropic size effects and additional interaction factors near the eutectic point, such as the proposed hydrogen bonding and/or the aforementioned weakening of the constant enthalpy of fusion assumption near the eutectic temperature.

For all cases examined in this study, the size-dependent model predicts phase equilibria better than the classic model.

\begin{SCfigure*}[\sidecaptionrelwidth][t!]
\centering
\includegraphics[width = 11.4 cm]{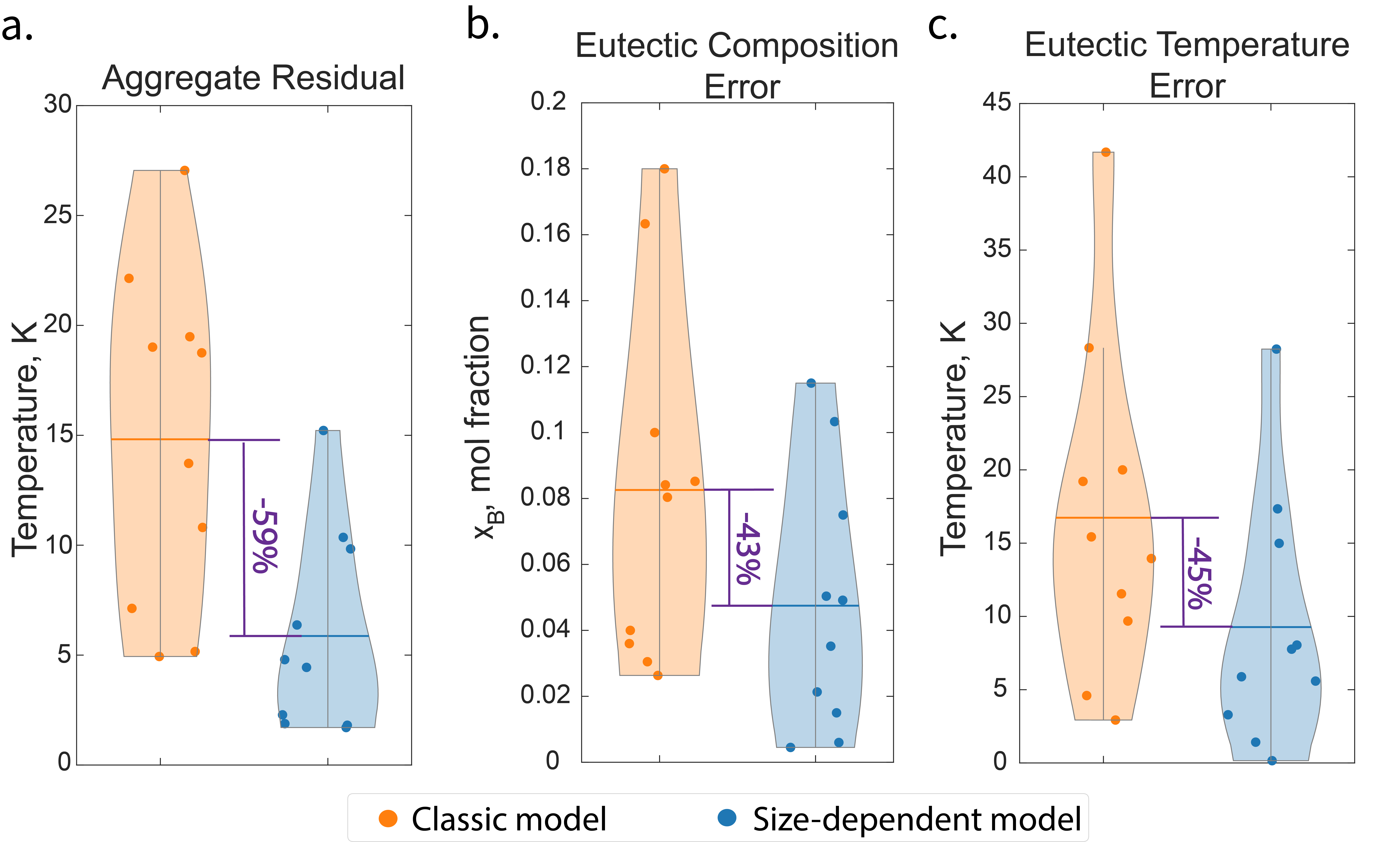}
\caption{\label{fig:Violins} Violin plots depicting data from Figure~\ref{fig:VIMRes}. Data is collapsed along $V_R$ to demonstrate average reduction for prediction error of (a) liquidus curves, (b) eutectic compositions, and (c) eutectic temperature as compared to experimental data. Solid lines indicate average values for each error. Error is significantly reduced in all cases, as indicated by purple brackets. Violin plots were generated using software from Bechtold \cite{bastian_bechtold_2021_4559847}.}
\end{SCfigure*}

To quantify this improvement in predictive power, Figure~\ref{fig:VIMRes}.a presents the aggregate residual temperature of the two models for each of the 10 solutions studied. This residual was calculated by numerically integrating the difference between predicted and experimental liquidus temperatures across the full range of mole fractions.

These data demonstrate that for all solutions, the size-dependent model has a lower total residual than the classic ideal model. On average, the size-dependent model resulted in a reduction in residual temperature of 59\% across all solutions sampled, as compared to the classic theory. Additionally, while the residual of the classic ideal model diverges, the size-dependent model performs similarly across all $V_R$, as should be expected if the proposed model better captures the physics of the system. Our model is demonstrated across the full range of mole fractions, and is therefore shown to be equally extensible to large solvents with small solutes.

Figure~\ref{fig:VIMRes}.b shows the error of the predicted eutectic composition, a solution property which is particularly laborious to obtain experimentally. It can again be seen that, while the predictive power of the size-dependent model is of varying accuracy, it consistently outperforms the classic model at predicting the eutectic composition. Figure~\ref{fig:VIMRes}.c shows an analogous plot for the error of predicted eutectic temperature, where the size-dependent model again performs better in all cases.

Interestingly, there does not appear to be a strong correlation with $V_R$ in predicted eutectic composition error reduction. The eutectic composition is impacted by the size-dependent entropy (as evidenced by the improvement seen here), but is also strongly dependent on melting temperature and enthalpy of fusion. It should be noted that many of the larger molecules investigated in this study possessed relatively high melting points (in excess of 400 K), which tend to drive their eutectic compositions to very low mole fractions of water. This may depress the magnitude of the error in both models (and particularly the classic model) and reduce the ability to compare eutectic composition predictions between solutions. We suggest that a more exhaustive study may reveal correlation of eutectic prediction error reduction with molar volume ratio, but the size-dependent model outperforms the classic ideal model regardless.

Finally, Figure~\ref{fig:Violins} recasts the data from Figure~\ref{fig:VIMRes} to better convey the absolute accuracy of each model. On average, the predicted size-dependent eutectic composition deviates 4.7 mol\% from experiment (3.6 mol\% closer than the classic model, for an average error reduction of 43\%); the predicted eutectic temperature deviates 9.7 K from experiment  (7.6 K closer than the classic model, for an average error reduction of 45\%); and the average aggregate residual in the liquid temperature deviates 5.9 K from experiment (9.9 K closer than the classic model, for an average error reduction of 59\%).

\section*{Discussion}

These results demonstrate that accounting for the simple effect of molecule size on mixing entropy can significantly improve our ability to predict aqueous organic solution thermodynamics, without significantly increasing theoretical complexity. We show this for 10 non-ionic aqueous organic solutions, and demonstrate that this theory retains its predictive power across disparate solute volumes even as the classical ideal solution model diverges. 

Like the classic model, this size-dependent theory relies only on properties of the pure components (here, molar volume, enthalpy of fusion, and melting temperature), each assumed constant with composition and temperature. As such, the model loses accuracy as it approaches the eutectic, where these assumptions of constancy become weakest, and where high viscosity may furthermore increase solid-state miscibility \cite{koop2000} and weaken the assumption of a pure crystalline phase. The theory also assumes that the excess volumes of components in solution are negligible. For aqueous organic solutions, this assumption is typically strong (with excess volumes generally remaining $<$1\% \cite{Powell-Palm2024}), and we confirm in Supplementary Note 1 that plausible changes in molar volume due to thermal expansion/contraction, excess mixing volume, and use of room temperature liquid vs. solid pure component molar volumes  do not significantly affect the accuracy of the model. However, we anticipate that extension of this model to highly interacting molecules will require correction for larger excess volumes. 

 In particular, correction for excess volumes may enable translation of this theory to aqueous solutions containing dissociated inorganic ions, which, given their strong hydration behaviors, confound straightforward interpretation of solute volume. Additionally, large ionic solvent shells may substantially affect the configurational nature of the solution (as well as changing the effective solute size) \cite{Martin2023}. Surmounting these barriers to a size-dependent understanding of ionic aqueous solutions could lead to improved thermodynamic understanding in several domains, including aqueous electrochemistry \cite{Liang2023}, deep eutectic solvents \cite{Smith2014}, desalination \cite{Najim2022}, and the planetary science of icy worlds \cite{Vance2021}. 

Furthermore, extension of this theory to multicomponent systems may enable predictive synthesis of arbitrarily complex aqueous organic solutions with desired phase equilibria, such as NADES materials for industrial chemistry \cite{Martins2019}, next-generation anti-freeze solutions for aircraft de-icing, and cryoprotectant media for organ and tissue preservation\cite{Higgins2} In each potential application, physical understanding of the thermodynamic factors driving solution solubility and melting point depression limit engineering progress and curtail the available solution design space. For example, increasingly accurate \textit{a priori }prediction of aqueous phase equilibria can accelerate the pressing search for new cryoprotectants\cite{Ahmadkhani2024, Higgins2, HigginsCryobio}, which must balance the capacity of organic solutes to protect against ice formation (i.e. balance their melting point depression and eutectic temperatures) against their stability in solution (i.e. solubility) and their relative toxicity.

In sum, this study suggests a dominance of size-dependent entropic contributions over the thermodynamics of aqueous organic solutions that is not broadly captured by existing solution theory literature. In particular, widely held conceptions of non-ideal solution thermodynamics either \textit{cannot} conveniently isolate these contributions from those originating from molecular interactions (as in activity models, wherein deviation from ideality is generally captured by a single activity coefficient), or \textit{do not} do so (as in the regular solution model, where deviations from ideality are generally captured by interaction parameters). We suggest that this is likely a product of simple historical happenstance— Hildebrand, Guggenheim, and the other early contributors to the theory of regular or non-ideal solutions, all of whom recognized the effect of unequal molecule sizes on the entropy of mixing, happened to study classes of solutions for which this effect proved largely negligible \cite{HildebrandBook}. We further note that the intriguing recent descriptive models of Martin and colleagues, which capture \textit{de facto} size-effects using additional solution-specific parameters related to solvation, have achieved remarkable agreement with experiments\cite{Martin2023}.

Finally, given the function of the present model as a "drop-in" alternative to the classic ideal, we anticipate that future work could readily extend it into new formulations of the many models that stem from the ideal model, establishing size-dependent regular solution models, size-dependent activity models, etc. These extensions may reduce the remaining gap between this ideal model and experimental observation, and may, by isolating purely size-dependent entropic effects, better capture the specific role of molecular interactions in liquid solutions.

 In any case, we here observe that the role of molecule size is significant in many aqueous solutions of relevance to biology, medicine, and organic chemistry, and we anticipate that further interrogation of entropic size effects in these systems will yield even greater insight into their thermodynamics. 

\acknow{This work was supported by funding from the NSF Engineering Research Center for Advanced Technologies for Preservation of Biological Systems (ATP-Bio), NSF EEC No. 1941543.}

\showacknow{} 

\bibsplit[8]

\bibliography{VIM}

\begin{thebibliography}{10}

\bibitem{Martins2019}
MA Martins, SP Pinho, JA Coutinho, {Insights into the Nature of Eutectic and Deep Eutectic Mixtures}.
\newblock {\em\protect\JournalTitle{Journal of Solution Chemistry}} \textbf{48}, 962--982 (2019).

\bibitem{Murray2022}
KA Murray, MI Gibson, {Chemical approaches to cryopreservation}.
\newblock {\em\protect\JournalTitle{Nature Reviews Chemistry}} \textbf{6}, 579--593 (2022).

\bibitem{Paiva2013}
A Paiva, et~al., Natural deep eutectic solvents – solvents for the 21st century.
\newblock {\em\protect\JournalTitle{ACS Sustainable Chemistry \& Engineering}} \textbf{2}, 1063--1071 (2014).

\bibitem{Garbett2012}
NC Garbett, JB Chaires, Thermodynamic studies for drug design and screening.
\newblock {\em\protect\JournalTitle{Expert Opinion on Drug Discovery}} \textbf{7}, 299--314 (2012) PMID: 22458502.

\bibitem{Klebe2015}
G Klebe, {Applying thermodynamic profiling in lead finding and optimization}.
\newblock {\em\protect\JournalTitle{Nature Reviews Drug Discovery}} \textbf{14}, 95--110 (2015).

\bibitem{oates1969ideal}
W Oates, Ideal solutions.
\newblock {\em\protect\JournalTitle{Journal of Chemical Education}} \textbf{46}, 501 (1969).

\bibitem{Hildebrand1927}
JH Hildebrand, {A Quantitative Treatment of Deviations from Raoult's Law}.
\newblock {\em\protect\JournalTitle{Proceedings of the National Academy of Sciences}} \textbf{13}, 267--272 (1927).

\bibitem{HildebrandBook}
JH Hildebrand, JM Prausnitz, RLRL Scott, {\em Regular and related solutions : the solubility of gases, liquids, and solids}.
\newblock (Van Nostrand Reinhold Co.), (1970).

\bibitem{Gordon1968}
P Gordon, {\em Principles of Phase Diagrams in Materials Systems}.
\newblock (McGraw-Hill, New York), (1968).

\bibitem{robinson1949tables}
RA Robinson, RH Stokes, Tables of osmotic and activity coefficients of electrolytes in aqueous solution at 25 c.
\newblock {\em\protect\JournalTitle{Transactions of the Faraday Society}} \textbf{45}, 612--624 (1949).

\bibitem{arrhenius1912theories}
S Arrhenius, {\em Theories of Solutions}, Mrs. Hepsa Ely Silliman memorial lectures.
\newblock (Yale University Press), (1912).

\bibitem{Martin2023}
JD Martin, AM Shipman, {Deep Dive into Eutectics: On the Origin of Deep and Elevated Eutectics}.
\newblock {\em\protect\JournalTitle{Journal of The Electrochemical Society}} \textbf{170}, 066508 (2023).

\bibitem{Flory}
P Flory, {Thermodynamics of high polymer solutions}.
\newblock {\em\protect\JournalTitle{Journal of Chemical Physics}} \textbf{22}, 415--426 (1942).

\bibitem{Huggins1942}
ML Huggins, {Theory of Solutions of High Polymers}.
\newblock {\em\protect\JournalTitle{Journal of the American Chemical Society}} \textbf{64}, 1712--1719 (1942).

\bibitem{Powell-Palm2024}
MJ Powell-Palm, H Smith, MM Fahad, {An entropic theory of homogeneous ice nucleation in non-ionic aqueous solutions}.
\newblock {\em\protect\JournalTitle{Journal of Chemical Physics}} \textbf{160} (2024).

\bibitem{Hildebrand1948}
JH Hildebrand, {A critique of the thoery of solubility of non-electrolytes}.
\newblock {\em\protect\JournalTitle{Symposium on Thermodynamics and Molecular Structure of Solutions, American Chemical Society}} pp. 37--45 (1948).

\bibitem{coutinho1995}
JA Coutinho, SI Andersen, EH Stenby, Evaluation of activity coefficient models in prediction of alkane solid-liquid equilibria.
\newblock {\em\protect\JournalTitle{Fluid Phase Equilibria}} \textbf{103}, 23--39 (1995).

\bibitem{Hansen1988}
JH Hansen, A Fredenslund, KS Pedersen, HP Rønningsen, A thermodynamic model for predicting wax formation in crude oils.
\newblock {\em\protect\JournalTitle{AIChE Journal}} \textbf{34}, 1937--1942 (1988).

\bibitem{Hubel2021}
K Hornberger, R Li, ARC Duarte, A Hubel, Natural deep eutectic systems for nature-inspired cryopreservation of cells.
\newblock {\em\protect\JournalTitle{AIChE Journal}} \textbf{67}, e17085 (2021).

\bibitem{Ahmadkhani2024}
N Ahmadkhani, JD Benson, A Eroglu, AZ Higgins, High throughput method for simultaneous screening of membrane permeability and toxicity for discovery of new cryoprotective agents.
\newblock {\em\protect\JournalTitle{bioRxiv}} (2024).

\bibitem{porter}
D Porter, K Easterling, M Sherif, {\em Phase Transformations in Metals and Alloys (4th ed.)}.
\newblock (CRC Press), (2021).

\bibitem{Murray2008}
BJ Murray, AK Bertram, Inhibition of solute crystallisation in aqueous h+nh4+so42–h2o droplets.
\newblock {\em\protect\JournalTitle{Phys. Chem. Chem. Phys.}} \textbf{10}, 3287--3301 (2008).

\bibitem{Hudait2014}
A Hudait, V Molinero, Ice crystallization in ultrafine water–salt aerosols: Nucleation, ice-solution equilibrium, and internal structure.
\newblock {\em\protect\JournalTitle{Journal of the American Chemical Society}} \textbf{136}, 8081--8093 (2014) PMID: 24820354.

\bibitem{Sangster1999}
J Sangster, {Phase Diagrams and Thermodynamic Properties of Binary Systems of Drugs}.
\newblock {\em\protect\JournalTitle{Journal of Physical and Chemical Reference Data}} \textbf{28}, 889--930 (1999).

\bibitem{Babkina2010}
TS Babkina, AV Kuznetsov, {Phase equilibria in binary subsystems of urea-biuret-water system}.
\newblock {\em\protect\JournalTitle{Journal of Thermal Analysis and Calorimetry}} \textbf{101}, 33--40 (2010).

\bibitem{Shnidman}
L Shnidman, AA Sunier, {The solubility of urea in water}.
\newblock {\em\protect\JournalTitle{Journal of Physical Chemistry}} (1932).

\bibitem{Ott1972}
JB Ott, JR Goates, JD Lamb, {Solid-liquid phase equilibria in water + ethylene glycol}.
\newblock {\em\protect\JournalTitle{The Journal of Chemical Thermodynamics}} \textbf{4}, 123--126 (1972).

\bibitem{Beyer2008}
KD Beyer, K Friesen, JR Bothe, B Palet, Phase diagrams and water activities of aqueous dicarboxylic acid systems of atmospheric importance.
\newblock {\em\protect\JournalTitle{The Journal of Physical Chemistry A}} \textbf{112}, 11704--11713 (2008).

\bibitem{Lane1925}
LB Lane, {Freezing Points of Glycerol and its Aqueous Solutions}.
\newblock {\em\protect\JournalTitle{Industrial and Engineering Chemistry}} \textbf{17}, 924 (1925).

\bibitem{Young1952}
BFE Young, FT Jones, HJ Lewis, {d-FRUCTOSE-WATER phase diagram}.
\newblock {\em\protect\JournalTitle{Journal of Physical Chemistry}} \textbf{222}, 1093--1096 (1952).

\bibitem{Young1922}
FE Young, {d-GLUCOSE-WATER PHASE DIAGRAM}.
\newblock {\em\protect\JournalTitle{Journal of Physcial Chemistry}} \textbf{61}, 616--619 (1957).

\bibitem{Siniti1999}
M Siniti, S Jabrane, JM L{\'{e}}toff{\'{e}}, {Study of the respective binary phase diagrams of sorbitol with mannitol, maltitol and water}.
\newblock {\em\protect\JournalTitle{Thermochimica Acta}} \textbf{325}, 171--180 (1999).

\bibitem{Young1949}
FE Young, FT Jones, {HYDRATES The Sucrose-Water phase diagram}.
\newblock {\em\protect\JournalTitle{Journal of Physical and Colloid Chemistry}} pp. 1334--1350 (1949).

\bibitem{Stephens}
H Stephen, T Stephen, {\em Solubilities of Inorganic and Organic Compounds}.
\newblock (MacMillan Co., New York), (1963).

\bibitem{bastian_bechtold_2021_4559847}
B Bechtold, P Fletcher, seamusholden, S Gorur-Shandilya, bastibe/violinplot-matlab: A good starting point (2021).

\bibitem{koop2000}
T Koop, A Kapilashrami, LT Molina, MJ Molina, Phase transitions of sea-salt/water mixtures at low temperatures: Implications for ozone chemistry in the polar marine boundary layer.
\newblock {\em\protect\JournalTitle{Journal of Geophysical Research: Atmospheres}} \textbf{105}, 26393--26402 (2000).

\bibitem{Liang2023}
Y Liang, Y Yao, {Designing modern aqueous batteries}.
\newblock {\em\protect\JournalTitle{Nature Reviews Materials}} \textbf{8}, 109--122 (2023).

\bibitem{Smith2014}
EL Smith, AP Abbott, KS Ryder, Deep eutectic solvents (dess) and their applications.
\newblock {\em\protect\JournalTitle{Chemical Reviews}} \textbf{114}, 11060--11082 (2014) PMID: 25300631.

\bibitem{Najim2022}
A Najim, {A review of advances in freeze desalination and future prospects}.
\newblock {\em\protect\JournalTitle{npj Clean Water}} \textbf{5} (2022).

\bibitem{Vance2021}
SD Vance, B Journaux, M Hesse, G Steinbrügge, The salty secrets of icy ocean worlds.
\newblock {\em\protect\JournalTitle{Journal of Geophysical Research: Planets}} \textbf{126}, e2020JE006736 (2021) e2020JE006736 2020JE006736.

\bibitem{Higgins2}
RM Warner, et~al., Rapid quantification of multi-cryoprotectant toxicity using an automated liquid handling method.
\newblock {\em\protect\JournalTitle{Cryobiology}} \textbf{98}, 219--232 (2021).

\bibitem{HigginsCryobio}
RM Warner, KS Brown, JD Benson, A Eroglu, AZ Higgins, Multiple cryoprotectant toxicity model for vitrification solution optimization.
\newblock {\em\protect\JournalTitle{Cryobiology}} \textbf{108}, 1--9 (2022).

\end{thebibliography}
\end{document}